\documentclass[11pt]{article}
\usepackage[dvips]{graphicx}
\usepackage{amsfonts,amssymb}
\usepackage{latexsym}
\usepackage{fancyheadings}
\usepackage{graphicx} 

\def\BibTeX{{\rm B\kern-.05em{\sc i\kern-.025em b}\kern-.08em
    T\kern-.1667em\lower.7ex\hbox{E}\kern-.125emX}}

\begin{document}
\baselineskip=18.6pt

\title{ Potential Links: Gravity and Nuclear Force}

\author{Shantilal G. Goradia\\ email: {\em Shantilalg}@juno.com}    
       
\maketitle

\abstract{The gravitational inverse square law is a macroscopic
approximation. I suggest that it should be modified for small
particles to use the surface-to-surface separation of particles rather
than the center-to-center separations. For small particles at
macroscopic separations, the ratio between the center-to-center
distance D and surface-to-surface distance d, D/d, approaches
unity. At microscopic separations, this ratio grows very large. Here I
apply this ratio to several microscopic situations and derive the
nuclear coupling constants. I will then present a model of a
gluon/graviton transformation to justify my surface originating
modification.} 

\section{Introduction}
Newtonian gravity encounters issues for microscopic dimensions.
As the sizes of two adjoining identical particles of uniform density tend to
zero, the numerator of the force equation ($F=Gm_1m_2/r^2$) falls off as $r^6$.
Since the denominator falls off as $r^2$, the force goes to zero in the limit 
of small particles with microscopic separations. Newtonian gravity in this
form, therefore, cannot explain the nuclear binding force.

If we use the surface-to-surface 
separation between these particles to quantify the gravitational attraction 
instead of the center-to-center separation, we find that the force between 
these microscopic particles is the same as before in the limit of large 
separations relative to the particle radii. At small separations relative to 
the particle radii the force between these same particles grows much larger 
than classical gravity. 

\section{Modification of the inverse square law}

For two coupled nucleons, I choose the Planck length
$L=\sqrt{(Gh/c^3)}$ as the surface separation, as it is the minimum possible 
spatial distance that makes any sense in physics. Assuming zero separation
distance would imply that the two particles are joined to form one particle,
losing their distinction as separate particles. The diameter of a nucleon is 
about $1$ fm ($10^{-15}$ meters). The Newtonian force
is then $F_N=Gm^2/D^2$, where $D$ is the center-to-center distance, $\sim$ 
$1$ fm. If we select the surface-to-surface separation instead, the force
would become $F_P=Gm^2/d^2$, with $d=L=10^{-20}$ fm. The ratio of these two 
forces is $D^2/d^2=10^{40}$, which is also the strength of the nuclear force
relative to gravitation. Strictly speaking, the strong force is not purely
short range decreasing to a precise zero beyond a boundary, as illustrated 
by Rutherford's scattering experiments, which showed
effects from the strong force at separations of at least $10$ fm [1].
As the nucleons are separated, $D/d$ shrinks, and $F_P$ rapidly approaches 
$F_N$ (Fig. 1). At $1000$ fm (less than the radius of an atom) the modified law
yields the same results as standard Newtonian gravitation.  Einstein  also tried to  explain nuclear force in terms of gravity [2]. 

For a coupled quark-lepton pair, the center separation can be taken
to be $\sim$ $10^{-3}$ fm. If we modify Newton's equation as above, we find 
that the ratio between the standard and modified force is $10^{34}$.
This is the relative strength of the weak nuclear force compared to 
gravitation. The weak nuclear force diminishes to standard Newtonian gravity
at a distance of $1$ fm, the diameter of the nucleon (Fig. 1).

\begin{figure}[t]
\centerline{\resizebox{4in}{2.7in}{\includegraphics{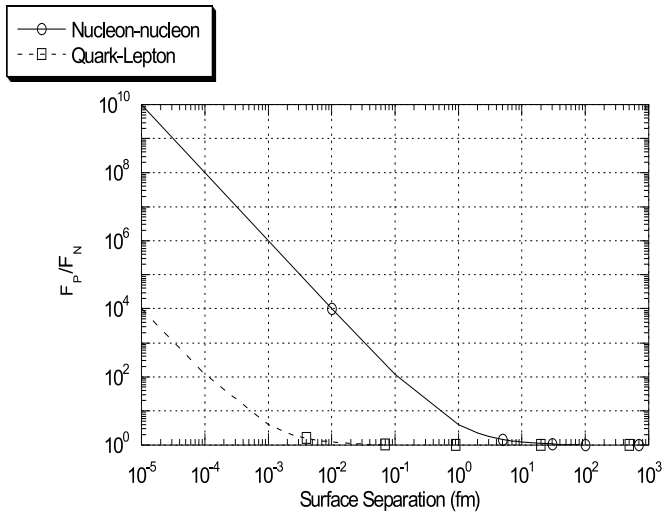}}}
\vspace{-0.1in}
\caption{Ratio of modified force to Newtonian gravitation
as a function of surface separation for nucleon-nucleon and quark-lepton
interactions. The ratio approaches unity at large surface separations in
both cases. Also, for both interactions the ratio becomes quite large
for short separations, reaching $10^{40}$ for the nucleon-nucleon interaction 
and $10^{34}$ for the quark-lepton interaction in the Planck length 
separation limit of $10^{-20}$ fm.} \label{fig:2}
\end{figure}
\vspace{-0.1in}

\section{Links to Current Physics}
    Gluons create an effect that is approximately the classical
equivalent of parallel springs connecting quarks: greater separations
lead to grater attractive forces.  If the hadron is considered to be a 
spherical ``sea'' of gluons surrounded by quarks, gluons may transform
to gravitons as they "evaporate" out of the gluon-sea or gravitons may
"condense" into gluons at this surface. As the virtual gluons
"evaporate", they can either become free gluons, or possibly pairs of
spin 1 gluons could transform into spin 2 gravitons. These
radiation/absorption processes would create 1/r potentials radiating
in all directions with their origins at the surface of the hadrons,
not at the center of the hadrons. This would create a very strong
force between the surfaces of closely packed particles, while being
indistinguishable from gravity at large distances.  Gravity between two spheres one fm
in diameter and a Planck length apart, when hypothesized my way yields
the strength of the nuclear force relative to gravity exactly and consistently. If the above transformation takes
Planck time, there will be a Planck length separation between the
quark-gluon plasma and the gravitational interaction. This would be consistent with general relativity and quantum mechanics; for this surface-originating gravity, the Compton wavelength would equal the gravitational radius, incorporating the uncertainty principle and general relativity [3]. This picture is compatible with the prevailing view that the nuclear force is a secondary effect of the color force. In this model, the hadron is treated as a geometrically spherical object. This is not unprecedented; de-Sitter's hadron model is also spherical [3]. If hadrons are not exactly spherical, the exact force might be affected, but not its order of magnitude. 
     
The existence of the gluon-graviton transformation at the surface of
the hadron presents the intriguing possibility that few, if any,
gravitons are found inside hadrons. The nuclear force is a secondary
effect of the color force. I am proposing that nuclear force is strong
gravity characterized by the above transformation at the surface of
the hadron. Combining these ideas, it can be seen that gravitons
should not be found within the hadron, as implicitly predicted
previously with more details [4]. Gravitons from outside the hadron should generally
transform back to gluons and many gluons would transform to gravitons
as they leave the hadron. This transformation brings all attributes of
the color force (mass, spin and color charges) into gravitation
without any significant modification of general relativity other than the Planck time described above.
     
Einstein, in a paper written in 1919, attempted to demonstrate that
his gravitational fields play an important role in the structure and
stability of elementary particles. His hypothesis was not accepted
because of gravity's extreme weakness [3]. My hypothesis would negate
this argument, providing a more consistent picture and supporting
Einstein's insight. In this case, Einstein's fields do play an
important part in particle structure; their unexpected strength leads
to identification as a qualitatively different force, i.e. the
residual strong force. 
     
The classical Einstein equation may be viewed as an equation for a
self-interacting spin-2 field in Minkowski space as concluded by
R. M. Wald. [5].  Combining Wald's statement with the above
gluon/graviton transformation and the proposed 1/r propagation could
mathematically link the color forces to general relativity,
reinforcing the prevailing view. In layman's terms: The leakage of the
mixed color forces is potentially the white color "evaporating" from
the surfaces of hadrons, ultimately perceived as gravity i.e., the
carriers of color force below Planck scale are transformed into
carriers of gravity above Planck scale as if the leaking parallel
springs weaken and line up in series. While the surface is not
precisely defined because of uncertainty, it is qualitatively
intuitive. If hadrons are not exactly spherical, the exact force might
be affected, but not its order of magnitude.

\section{Prediction}
My theory provides a consistent, intuitive and simplistic, but
mathematical explanation of the relative values of the strong, weak
and gravitational coupling constants, something no other single theory
has done. If, as has been suggested, G is decreasing with time, the
Planck length shortens, affecting the values of coupling constants,
past nuclear reaction rates and the accuracy of
radio-dating. Experimentally, my theory can be explored through close
investigation of the nuclear force at $\approx$~1-10 fm. Additionally, my
prediction that gravitons do not exist in hadrons may resolve the
difficulties string theory has in incorporating gravity and color
force at the same time.  
     
\section{Conclusion}
There is a potential link between gravity and nuclear force.
The nuclear coupling constants, therefore, are expressible as the
squares of the sum of the diameters of the involved particles, expressed in Planck lengths. 
     
\section*{Acknowledgment}
I thank Dr. Weber (http://www.nd.edu/~fweber) and Dr. Christopher
Kolda for their earlier comments.


\begin{thebibliography}{99}
\bibitem{1}Elton, L. R. B, {\it Introductory Nuclear Theory}, 17 (1966).
\bibitem{2}Cao, T. Y, {\it The conceptual foundation of quantum field theory}, 85 (1999).
\bibitem{3}Shrivastava, S. K., {\it Aspects of Gravitational
Interactions}, 90 (1998).
\bibitem{4}Goradia, S. G., {\it physics/0110001 }.
\bibitem{5}Wald, R. M., {\it General Relativity}, 383 (1984).
\end{thebibliography}
\end{document}